\documentclass[
prl,
reprint,
groupedaddress
]{revtex4-1}

\usepackage{amsmath,amssymb} 
\usepackage{amsmath} 
\usepackage{amsfonts} 
\usepackage{bm} 
\usepackage[dvipdfm,colorlinks=true,citecolor=blue,linkcolor=blue,filecolor=blue,urlcolor=blue]{hyperref} 
\usepackage[pdftex]{graphicx} 
\usepackage{braket} 
\usepackage{ulem}

\begin{document}

\title{Incommensurate quantum magnet based on $4f$-electron in a zigzag spin-1/2 chain of YbCuS$_{2}$}

\author{Takahiro Onimaru}
\email{onimaru@hiroshima-u.ac.jp}
\author{Yudai Ohmagari}
\author{Soichiro Mizutani}
\author{Rikako Yamamoto}
\altaffiliation[Present address: ]{Max Planck Institute for Chemical Physics of Solids, 01187 Dresden, Germany}
\author{Hikaru Kaneshima}
\author{Chikako Moriyoshi}
 \affiliation{
Department of Quantum Matter, Graduate School of Advanced Science and Engineering, Hiroshima University, Higashi-Hiroshima, Hiroshima 739-8530, Japan}
\author{Devashibhai T. Adroja}
\author{Dmitry Khyalyavin}
\author{Pascal Manuel}
 \affiliation{%
ISIS Facility, Rutherford Appleton Laboratory, Chilton, Oxon OX11 0QX, United Kingdom}
\author{Hidehiro Saito}
\author{Chisa Hotta}
 \affiliation{
Department of Basic Science, University of Tokyo, Meguro-ku, Tokyo 153-8902, Japan}



\date{\today}

\begin{abstract}
We performed high-resolution powder neutron diffraction experiments and discovered an elliptic helical incommensurate magnetic structure in the semiconducting rare-earth magnet YbCuS$_2$, featuring effective spin-1/2 Yb$^{3+}$ ions that form a zigzag chain. Upon cooling the sample to 0.2 K, we observed very weak magnetic peaks indexed with an incommensurate propagation vector \textbf{\textit{k}} = [0, 0.305, 0] along the zigzag chain. The magnitude of the magnetic moment is at least one-third smaller than the expected value for the Yb$^{3+}$ Kramers doublet ground state. In an applied magnetic field, up-up-down magnetic order was observed at 7.5 T, characterized by diffraction peaks indexed with \textbf{\textit{k}} = [0, 1/3, 0] and substantial uniform magnetic components. These observations agree well with theoretical calculations based on the density matrix renormalization group for a zigzag spin-1/2 model with isotropic Heisenberg interactions and off-diagonal symmetric $\Gamma$-type exchange interactions derived from material parameters. The theory elucidates the quantum mechanical nature of the incommensurate magnetism as 
remnant off-diagonal spin correlations in a nematic dimer-singlet state. 
\end{abstract}

\maketitle

{\it Introduction.}
Long-wave modulation in magnetic structures is becoming increasingly important for 
exploring topological properties in materials. 
Helical and conical magnets exhibit one-dimensional (1D) spiral arrangements of spins, 
and the combinations of more than two spirals in two dimensions (2D) give rise to skyrmions \cite{Bogdanov1994,Muehlbauer2009,Nagaosa2013} 
that are topologically protected. 
Helical magnets with a single handedness can produce chiral magnetism, which is characterized by a lack of inversion symmetry,
and can form robust 1D structures that are sources of topological transport~\cite{Togawa2012,Laliena2020}, 
where the incommensurate magnetism is indeed observed~\cite{Kousaka2007}. 
In working with these rich variants of magnets, their large spin moments help establish the physical mechanisms of either 
the competition among neighboring and longer-range spin exchange interactions 
or Dzyaloshinskii-Moriya (DM) interactions at the classical level~\cite{Yi2009}. 
Most existing theories and experiments indeed rely on `classical' spins~\cite{Yi2009,Buhrandt2013}, 
underscoring the significant potential of discovering `quantum' incommensurate magnetism. 
To date, we find only a few numbers of quantum incommensurate long-range orders; TiO$X$ ($X$ $=$ Cl, Br) exhibiting incommensurate-to-commensurate transitions due to the frustration effect~\cite{Smaalen2005,Ruckamp2005}, 
and exotic spin amplitude waves observed in iron oxide spinels~\cite{Perversi2018}, though the origin remains unclear.
Here, we mean such quantum incommensurate magnets by elusive when they satisfy two conditions: First, they are not the spin-density wave in metals driven by neither nesting of Fermi surface nor coupled to lattices, but the quantum insulating magnets.
Second, the inversion symmetry is unbroken in the sense of excluding those driven by DM interactions in classical magnets.

\par
\begin{figure}[tbp]
\begin{center}
\includegraphics[width=8.5cm]{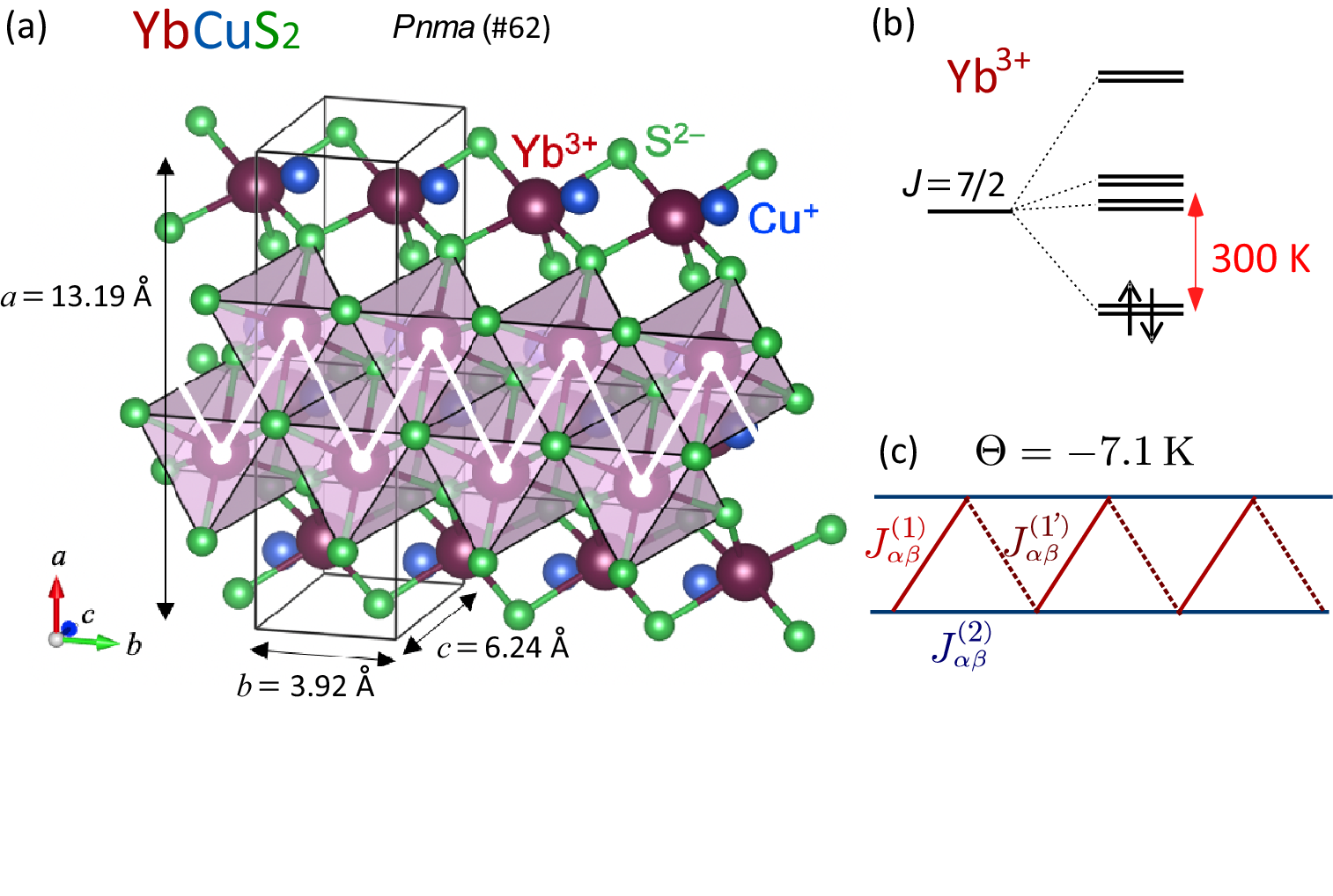}
\vspace{-10mm}
\caption{
(a) Orthorhombic crystal structure of YbCuS$_2$ with the space group of $Pnma$ (\#62) \cite{Strobel2007}.
The Yb$^{3+}$ ions are encapsulated within slightly distorted S$_6$ octahedra. The Yb site has a point group of $m$ ($C_{s}$).
The Yb ions form a zigzag chain along the $b$-axis. 
(b) Lowest Kramers doublet of Yb$^{3+}$ ion carrying effective spin-1/2.
(c) Zigzag chain model with anisotropic exchange interactions $J_{\alpha\beta}^{(\gamma)}$ among different spin components, $\alpha$, $\beta$ $=$ $x$, $y$, and $z$, 
for three bond directions, $\gamma$ $=$ 1, 1', and 2. 
The material has nearly isotropic $J_{xx}^{(\gamma)}\sim J_{yy}^{(\gamma)}\sim J_{zz}^{(\gamma)}$ common to $\gamma$ 
and the small off-diagonal ones called $\Gamma$ terms. 
}
\label{f1}
\end{center}
\end{figure}

In this study, we report another compelling case, 
the discovery of incommensurate magnetic structure of YbCuS$_2$, whose Yb$^{3+}$ 
ions carry effective spin-1/2 and form a zigzag chain (see Fig.~\ref{f1}(a,b))~\cite{Ohmagari2020}. 
One of the hallmarks of the spin-1/2 zigzag chain is the dimerized spin-singlet ground state in the antiferromagnetic Heisenberg model due to the interplay of quantum fluctuations and geometrical frustration~\cite{Okunishi2003,Okunishi2008,Hikihara2010}. 
However, YbCuS$_2$ is not such a simple Heisenberg magnet; because of the strong spin-orbit coupling of Yb ions, 
there arise an off-diagonal spin-exchange coupling called $\Gamma$ term~\cite{Saito2024} (Fig.~\ref{f1}(c)). 
Numerical and effective low energy theories show that a finite $\Gamma$ term 
converts the dimer-singlet phase to the coexisting nematic dimer singlet state, 
where the spin gap and the gapless modes are compatible. 
Intriguingly, despite its nonmagnetic nature, the state exhibits a robust diagonal and off-diagonal spin-spin correlation, 
indicative of quantum incommensurate helical magnetism. 
The present experimental measurement reveals the existence of $xy$-component helical modulation 
and the considerable reduction of magnetic moments, that is a rare example of quantum incommensurate magnetism 
amenable to laboratory observation. 
\par
\par
YbCuS$_{2}$ is a compensated semiconductor 
consisting of Yb$^{3+}$ (4f$^{13}$), Cu$^{+}$ (3d$^{10}$), and S$^{2-}$ ions~\cite{Ohmagari2020}, 
crystallizing in an orthorhombic crystal structure of a space group of $Pnma$~\cite{Strobel2007}. 
Due to the edge shared structure of the surrounding S$_{6}$ octahedra, 
the Yb$^{3+}$ ions form a zigzag chain along the $b$-axis, 
and its lowest energy doublet carry effective spin 1/2 (Fig.~\ref{f1}(b))~\cite{Lea1962}, 
interacting with each other through the superexchange interactions mediated by the two S$^{2-}$ anions. 
The material undergoes a first-order phase transition at $T_{\rm o}$ $=$ 0.95 K, 
where discontinuous quantities of the magnetization and the magnetic entropy satisfy the Claudius-Crapeiron relation~\cite{Ohmagari2020-2,Ohmagari2020}. 
The paramagnetic Curie temperature is evaluated as 
$\Theta$~$=$~$z4J_{\text{eff}}(J_{\text{eff}}+1)/(g\mu_B)^2\lambda/3=-7.1$ K from 
$\lambda=-10.5$ mol/emu and $J_{\text{eff}}=1/2$ (see Supplemental Materials \cite{supplement}), compared to $\theta_{\rm p}$ $=$ $-$92 K obtained from the high temperature magnetic susceptibility data~\cite{Ohmagari2020}.
In applying a magnetic field, the phase below $T_{\rm o}$ is found to be robust up to 4 T, 
and at $B\ge$ 4 T, the 1/3 plateau phase is expected from the isothermal magnetization
in terms of the magnetic moment carried by the ground state doublet. 
Recently, in a $^{63/65}$Cu nuclear quadrupole resonance (NQR) measurement using a powdered sample, 
a broadening of the spectrum was observed below $T_{\rm o}$, 
implying incommensurate modulation in the ordered state \cite{Hori2023}. 
Its spin-lattice relaxation rate $T_{1}^{-1}$ shows a $T$-linear behavior, 
indicating that the gapless spin excitation is present~\cite{Hori2023}. 
The present work clarifies the magnetic structure of this intriguing low-temperature phase. 
\par
{\it Experimental.}
Polycrystalline YbCuS$_{2}$ samples are synthesized using Yb, Cu, and S elements by utilizing the melt-growth technique 
(see our previous report in Ref. [\onlinecite{Ohmagari2020}]). 
The crystal structure of YbCuS$_{2}$ using the X-ray diffraction 
first reported on a powdered sample was orthorhombic with a space group of $P2_{1}2_{1}2_{1}$ \cite{Gulay2005}, 
similar to that of YCuS$_{2}$ \cite{Gulay2005-2}, 
and later as $Pnma$ for a single crystal~\cite{Strobel2007}. 
We conducted a powder X-ray diffraction measurement with the Debye-Scherrer geometry 
at beamline BL02B2 of SPring-8 in Japan \cite{Kawaguchi2017}, and based on the Rietveld analysis Jana program~\cite{Petricek2014}, 
confirmed the orthorhombic $Pnma$ structure shown in Fig. \ref{f1}(a) (see Supplemental Materials \cite{supplement}.). 
\begin{figure}[t]
\centering
\includegraphics[width=8.5cm]{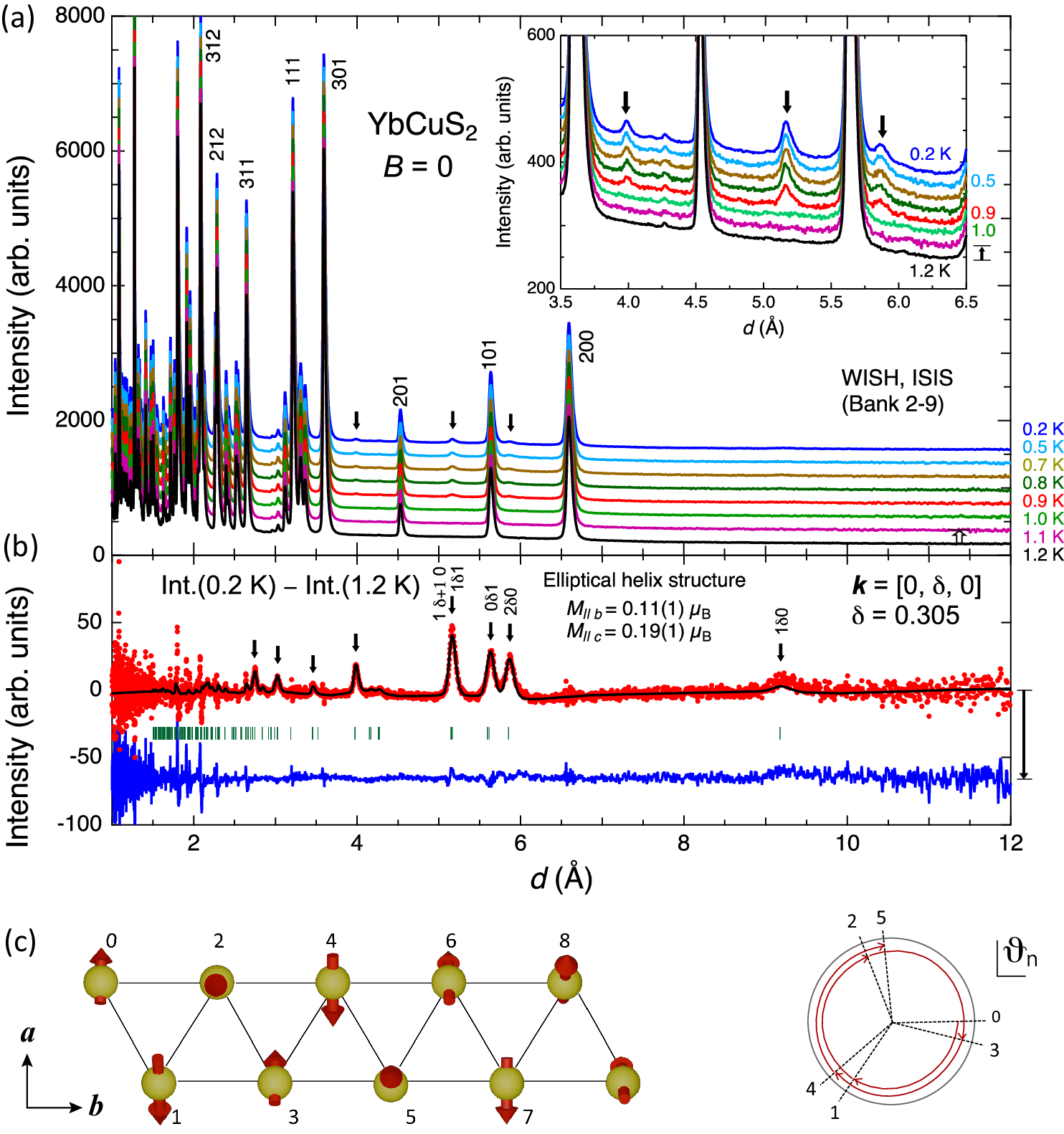}
\caption{
(a) Powder neutron diffraction patterns of YbCuS$_2$ measured in the temperature range from 0.2 K to 1.2 K. 
The inset displays the extended data between $d$ $=$ 3.5 and 6.5 {\AA}.
The arrows indicate the superlattice reflections observed below 0.9 K. 
(b) Red data points represent the difference between the diffraction patterns of 0.2 K and 1.2 K. 
The solid line is obtained by calculating an elliptic helical incommensurate magnetic structure. 
The blue line indicates the difference between the measured data and the calculation. 
The green bars represent the $d$ values expected for the magnetic peaks.
(c) Proposed magnetic structure with a propagation vector of 
\textbf{\textit k}$_{\text{ic}}=$ [0, 0.305, 0] (for unit cell), 
where the moments rotate in the $ac$-plane clockwise
with the phase factor $\vartheta_n=-0.695\pi n$ ($n$: integer) in a unit of Yb ion. 
}
\label{f2}
\end{figure}
\par
Powder neutron diffraction measurements are conducted with the high-resolution long-wavelength time-of-flight diffractometer WISH installed at the ISIS facility in the Rutherford Appleton Laboratory in the UK \cite{Chapon2011}. 
The scattering angle ranges from 10 to 175~deg., which corresponds to the $d$ spacing range between 0.7 and 50~{\AA}. 
A $^3$He-$^4$He dilution refrigerator was used to cool a 3.0-gram powdered sample sealed in a copper cell with $^4$He exchange gas 
down to the lowest temperature of 0.2 K, and a superconducting magnet generates magnetic fields up to 13 T. 
The neutron diffraction patterns are analyzed with the Rietveld analysis FullProf program 
\cite{RodrguezCarvajal1993} to evaluate the structural and magnetic contributions. 
\\
\begin{figure}
\centering
\includegraphics[width=8.5cm]{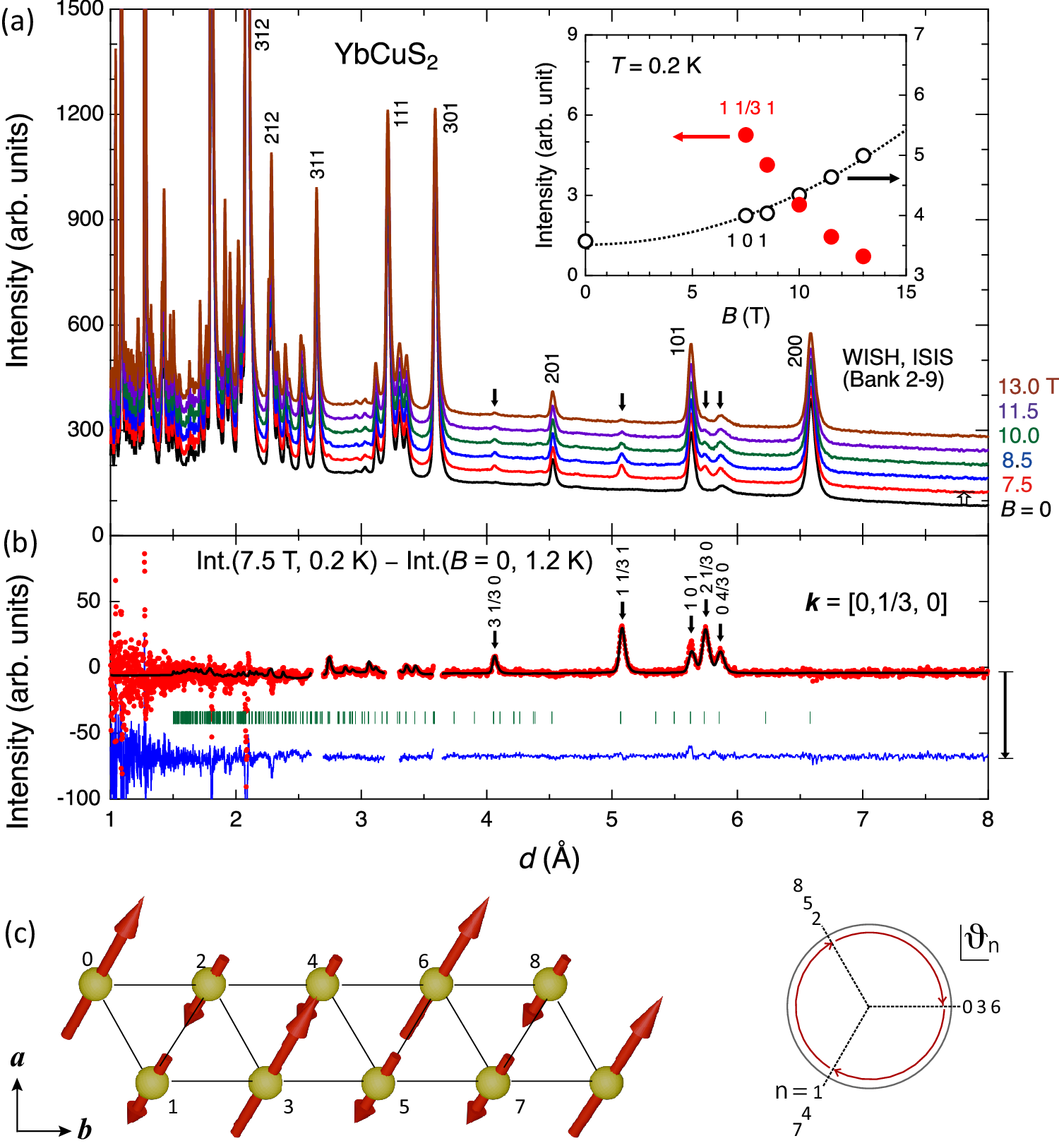}
\caption{
(a) Powder neutron diffraction patterns of YbCuS$_2$ measured at 0.2 K 
in the magnetic fields of $B$ $=$ 0, 7.5, 8.5, 10.0, 11.5, and 13.0 T. 
The inset shows the field dependent intensities of the Bragg peaks at 1 1$/$3 1 and 101.  
(b) Red data points represent the difference between the diffraction patterns of $(T,B)$= (0.2 K, 7.5 T) and (10 K, 0 T). 
The solid line is obtained by calculating the up-up-down magnetic structure. 
The blue line indicates the difference between the measured data and the calculation. 
The green bars are the expected $d$ values for the magnetic reflections.
(c) Proposed magnetic structure with a propagation vector of 
\textbf{\textit k} $=$ $\textbf{\textit k}_{1/3}$ $=$ [0, 1$/$3, 0] for a unit cell along the $b$-axis, 
rotating clockwise with the phase factor $\vartheta_n=-2\pi n/3$ in a unit of Yb ion. 
}
\label{f3}
\end{figure}
{\it Measurements.} 
Figure \ref{f2}(a) shows powder neutron diffraction patterns measured from 0.2 K to 1.2 K.
Nuclear peaks with integer $hkl$ indexes, such as 200, 101, and 201, are observed, 
and the intensity is independent of the temperature.
Weak superlattice peaks indicated by bold arrows are visible below 0.9 K, 
signifying the development of the magnetic order. 
In Fig. \ref{f2}(b), the difference between the diffraction below (0.2 K) and above (1.2 K) $T_{\rm o}$ are 
shown; some superlattice peaks become more visible, indexed with a propagation vector of 
\textbf{\textit k}$_{\text{ic}}$ $=$ [0, 0.305, 0], 
which corresponds to the incommensurate modulation in the Yb zigzag chain along the $b$-axis.
They are compared with the calculation assuming 
the elliptic helical incommensurate modulation of the Yb moments in the $ac$-plane
with propagation vector $\bm k_{\text{ic}}$ 
and with the components of $\mu_{x}$ $=$ 0.378(2) and $\mu_{z}$ $=$ 0.214(2) $\mu_{\rm B}$/Yb; 
the difference between the measured data and the calculation is shown with blue data points in Fig.~\ref{f2}(b), 
confirming that the intensity of the measured peaks and the $d$ values are well reproduced. 
It is worth noting that the observed moments are at least by a factor of three 
smaller than the expected full moment of the ground state doublet 1.3--1.4 $\mu_{\rm B}$/Yb. 
From these observations, we propose a magnetic structure in Fig. \ref{f2}(c), 
$\big(\mu_x\sin(\vartheta_n),0,\mu_z\cos(\vartheta_n)\big)$ with 
$\vartheta_n=-0.695\pi n$ rotating clockwise for Yb ions indexed by $n$ $=$ 0, 1, 2, $\cdots$. 
Other magnetic structure models are examined but are found to be less relevant (see Supplemental Materials \cite{supplement}.)
\par
Next, we show the powder neutron diffraction patterns 
in Fig.~\ref{f3}(a) measured at $T$ $=$ 0.2 K and $B$ up to 13.0 T. 
The field induced superlattice peaks are indicated by the arrows, 
indexed with a commensurate propagation vector of 
$\textbf{\textit k}_{1/3}$ $=$ [0, 1/3, 0] for a unit cell. 
The red data points in Fig. \ref{f3}(b) represent the difference between the diffraction patterns measured 
at $(T, B)$= (0.2 K, 7.5 T) and (10 K, 0 T), which are the up-up-down (UUD) and paramagnetic phases, respectively. 
Representative peaks that become visible include two components; 
the $hkl$ ones like 101, indexed with $\textbf{\textit k}$ $=$ 0, attributed to the uniform component of the Yb moments induced by the magnetic field, 
and the ones indexed with $\textbf{\textit k}_{1/3}$ due to the modulation along the $b$-axis. 
We fit the $B$ $=$ 7.5 T data by assuming both the uniform and $\textbf{\textit k}_{1/3}$ components, 
finding that the intensity of the peaks and the $d$ values are both reproduced 
with the magnetic components of $\mu_x = 0.432(8)$, $\mu_y = 0.472(8)$, and $\mu_z = 0.282(10)$ in unit of $\mu_{\rm B}$/Yb, 
where we can propose a magnetic structure of $\big(\mu_x$, $\mu_y$, $\mu_z\big)$$\cos(\vartheta_n)$ 
with $\vartheta_n=-2\pi n/3$ indexed by Yb ions for $n$ $=$ 0, 1, 2, $\cdots$ 
as shown in Fig.~\ref{f3}(c) 
(see Supplemental Materials \cite{supplement}.)
\par
\begin{figure}[t]
\centering
\includegraphics[width=8.5cm]{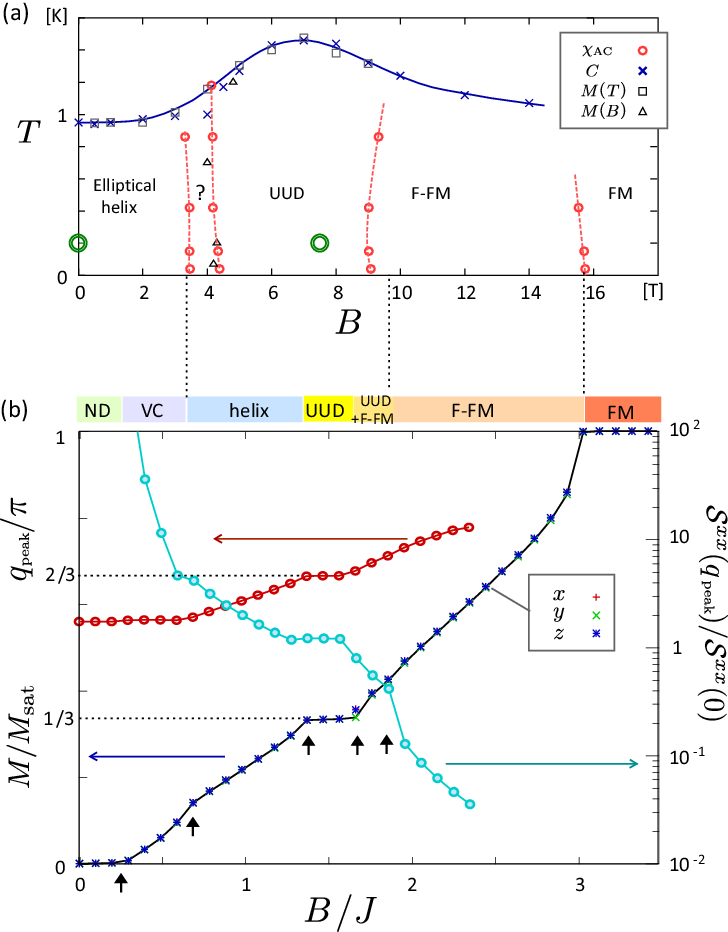}
\caption{
(a) Experimental $B$--$T$ phase diagram of YbCuS$_2$ obtained by AC magnetic susceptibility ($\chi_{\text{AC}}$), 
specific heat ($C$), DC magnetization ($M$) measurements~\cite{Ohmagari2020}. The measured points by means of the neutron diffraction at $T$ $=$ 0.2 K and in $B$ $=$ 0 and 7.5 T are shown with the (green) double circles.
(b) Theoretical ground state phase diagram of Eq.~(\ref{eq:ham}) obtained by DMRG with the $N=100$ system 
using the model parameters evaluated from the structural data of YbCuS$_2$~\cite{Saito2024}. 
Magnetization curve $M/M_{\text{sat}}$ for $\alpha$ $=$ $x$, $y$, $z$ field directions, 
peak position $q=q_{\text{peak}}$ of the structure factor 
${\cal S}^{\alpha\beta}(q)$, and ${\cal S}^{xx}(q_{\text{peak}})/{\cal S}^{xx}(0)$ for the $x$ field direction. 
We denote each phase as nematic dimer-singlet (ND), vector chiral (VC), up-up-down (UUD) plateau, and fluctuating-ferromagnet (F-FM). 
}
\label{f4}
\end{figure}

   
In Figure~\ref{f4}(a), the experimentally derived $B$--$T$ phase diagram was
obtained by analyzing the anomalies in the AC magnetic susceptibility, specific heat, and DC magnetization \cite{Ohmagari2020}.
The diagram reveals five different magnetic phases as the magnetic field increases.
The (green) double circles represent the measured points in the present neutron diffraction experiments at $T$ $=$ 0.2 K and in $B$ $=$ 0 and 7.5 T.
The transition from the incommensurate magnetic structure to the UUD structure is considered reasonable.

{\it Comparison with theory.} 
To understand the $B$--$T$ phase diagram and the evolution of the order parameters, we perform the density matrix renormalization (DMRG) calculation~\cite{White1993} on 
a quantum spin-1/2 zigzag chain model, which is derived microscopically based on the experimentally determined structures of YbCuS$_2$, 
given in the form, 
\begin{align}
{\cal H}= \sum_{i,j} \sum_{\alpha,\beta=x,y,z} J^{(\gamma)}_{\alpha\beta} S_i^\alpha S_j^\beta - {\bm B} \cdot \sum_{j} {\bm S_j}. 
\label{eq:ham}
\end{align}
Here, $J^{(\gamma)}$ is the $3\times 3$ matrix that represents the spin exchange interactions \cite{supplement}, and the second term is the contribution of the Zeeman effect.
We have dominant Heisenberg term, $J_{xx}\sim J_{yy}\sim J_{zz}$, 
and the small off-diagonal terms $J_{\alpha\beta}\sim J_{xx}/20$ qualitatively equivalent to the $\Gamma$ terms, 
whereas the DM interaction is strictly absent~\cite{Saito2024}. 
It is to be compared with the ground state phases of Eq.~(\ref{eq:ham}) in Fig.~\ref{f4}(b); they are determined 
based on the magnetization curve $M/M_{\text{sat}}$ ($M_{\text{sat}}$ is the full moment), 
peak position $q=q_{\text{peak}}$ of the structure factor of spin-spin correlation 
${\cal S}^{\alpha\beta}(q)=N^{-1}\sum_{j,j'}S_j^\alpha S_{j'}^\beta e^{iq(j-j')}$, 
and the amplitude of ${\cal S}^{xx}(q_{\text{peak}})/{\cal S}^{xx}(0)$. 
Here, ${\cal S}^{\alpha\beta}(q)$ has equal diagonal 
$\alpha\beta$ $=$ $xx$, $yy$, $zz$ components and the small off-diagonal ($\alpha\ne\beta$) ones, and their incommensurate oscillation periods are analyzed as $q_{\text{peak}}=0.568\pi$ \cite{supplement}.
Supposing that the experimental saturation field $B\sim 15.7$ T is equal to $B/J=3$, 
the average Heisenberg exchange coupling is given as $J\sim 3.5$ K, 
and accordingly, the spin gap is $0.35 J\sim 1.2$ K, which is close to $T_{\text{o}}$. 
This suggests that the transition at $T_{\text{o}}$ basically opens a spin gap.
It is noteworthy that the anomalies in $\chi_{\text{AC}}$ detected at around 3.5 T agree well 
with the endpoint of the nonmagnetic (nematic dimer-singlet (ND) and vector chiral (VC)) phases of the model. 
The boundaries with the UUD and F-FM at around 9 T in the experiment also agree with the theory. 
\par
Let us discuss their magnetic structure in more detail. 
As seen in the inset of Fig. \ref{f3}(a), the intensity of the 1 $1/3$ 1 reflection caused by the UUD structure decreases, 
while the 101 reflection increases with $\propto B^{2}$ as the magnetic field is increased up to 13.0 T. 
Because they show no anomaly at around 9 T, the switching of the order parameter can be ruled out. 
Instead, it may result from a kind of crossover from the modulated structure to the uniform one. 
Indeed, in the theoretical calculation, a fluctuating ferromagnetic (F-FM) state appears between 
the UUD phase and the forced FM phase \cite{Saito2024-3}; 
as shown in Fig.~\ref{f4}(b), 
the peak position $q_{\text{peak}}$ gradually changes from $2\pi/3$ as the state moves away from the pure UUD state. 
Namely, the magnetic field gradually alters the UUD alignment to the forced ferromagnetic order. 
A similar phase transition was also observed in a Yb-based semiconductor YbAgSe$_{2}$, 
which crystallizes in an orthorhombic structure with the space group of $P2_{1}2_{1}2_{1}$ 
\cite{JulienPouzol1977,Mizutani2022}.
\par
{\it Summary and discussion.} 
We performed powder neutron diffraction experiments to investigate the magnetic structure of the lowest temperature phase 
of the semiconductor YbCuS$_{2}$. 
On cooling the sample below the transition temperature of $T_{\rm o}=$ 0.95 K, 
we observed very weak magnetic peaks with an incommensurate propagation vector of 
\textbf{\textit k}$_{\text{ic}}$ = [0, 0.305, 0]. 
By analyzing the diffraction pattern, 
it was found that the Yb moments are confined to the $ac$ plane and are squeezed at least by one-third of magnitude to 
$\mu_{a}$ $=$ 0.38(2) and $\mu_{c}$ $=$ 0.22(2) $\mu_{\rm B}$/Yb, 
forming an elliptic helical modulation along the zigzag chain. 
In a magnetic field at 7.5 T, the distinct UUD structure known as 1/3-magnetization plateau state was observed, 
which gradually transforms to an F-FM phase at higher fields. 
\par
Theoretical phase diagram obtained by DMRG shows good agreement both in terms of the magnetic-field dependence 
and on the details of the magnetic structures. 
In particular, the zero-field phase is considered to be the ND phase~\cite{Saito2024-2}, 
having a finite spin gap and a gapless nematic excitation \cite{Saito2024-3}. 
In the Heisenberg zigzag spin chain, the pure dimer spin-singlet with a spin gap is established
\cite{Okunishi2003,Okunishi2008,Hikihara2010}, 
while the off-diagonal exchange coupling called $\Gamma$-term induces a gapless nematic mode 
without destroying the singlet. 
It is noteworthy that even in the pure dimer spin-singlet state, the diagonal ($S^\alpha_iS^\alpha_j$)
spin-spin correlation typical of the Tomonaga-Luttinger liquid survives, 
while the off-diagonal ones are strictly absent~\cite{Bursill1995,White1996}. 
In introducing the $\Gamma$-term, the $S^x_iS^y_j$ correlation develops 
and shows an incommensurate period close to the experimentally observed ones \cite{Saito2024-3}. 
This indicates that the incommensurate magnetism found in YbCuS$_2$ is of `quantum' origin. 
While the theory considers a purely 1D system, 
the material has a small but finite inter-zigzag chain interaction, possibly smaller by 
one order from the intra-chain $J\sim 3.5$ K in theory and 7.1 K in experiment~\cite{Ohmagari2020}. 
Once the ND phase is formed at $T_{\text{o}}$, 
the remaining interactions are inter-chain ones, 
which may drive the incommensurate correlation to a stable magnetic structure captured in the experiment. 
Still, the moments induced is  0.22--0.38 ${\mu}_{\rm B}$/Yb, much smaller than the full moment of
1.3 $\mu_{\rm B}$/Yb expected from the CEF ground state, 
which should be because the state is supported by the robust dimer-singlet. 
\par
Previously, the incommensurate magnetic orders in rare-earth-based intermetallic compounds were frequently observed due to the long-range Ruderman--Kittel--Kasuya--Yosida (RKKY) interaction mediated by the conduction electrons, which does not apply to the semiconducting state in YbCuS$_{2}$. 
This material has much in common to rare earth quantum magnets, previously discussed in relevance to the quantum spin liquid state, such as ytterbium 4f$^{13}$ triangular lattice compounds, YbMgGaO$_{4}$\cite{Li2015,Li2015-2,Li2016,Li2017,Bachus2020,Majumder2020,Rao2021,Dai2021,Xie2023} 
and $A$Yb$X_{2}$ ($A$ $=$ Na, Cs; $X$ $=$ O, S, and Se) \cite{Baenitz2018,Ranjith2019,Ranjith2019-2,Xing2019}. 
Their particular feature is the anisotropic exchange interaction induced by the large spin-orbit coupling of Yb ions and the crystalline electric field effect \cite{Rau2016,Lea1962}. 
In our case, the $\Gamma$-term allows the exotic quantum incommensurate magnetism possibly coexisting with the robust spin singlet phase. 
The observed incommensurate modulation of the Yb moments is consistent with the broadening of the NQR spectrum below $T_{\rm o}$ \cite{Hori2023}, which also reports the gapless excitation based on the $T$-linear behaviour of the $T_1^{-1}$. 
Because such gapless mode is well explained within the present framework as the signature of spin nematics, the material would also serve as a platform for studying an intriguing coexistence of spin singlet, spin nematics and the quantum incommensurate magnetism, which needs further clarifications in the family of rare earth materials $R$CuS$_2$ ($R=$ Dy, Ho, Er, Tm)~\cite{Ohmagari2020-2}.

{\it Acknowledgments.} 
The authors would like to thank H. Suzuki, Y. Shimura, K. Umeo, T. Takabatake, F. Hori, S. Kitagawa, K. Ishida, T. Ishizaki, H. Sagayama, and K. Ohoyama for the helpful discussion. 
The authors also thank Y. Shibata for the electron-probe microanalysis carried out at N-BARD, Hiroshima University. 
The authors gratefully acknowledge the technical staff at ISIS for technical support and assistance in this work.
This work was financially supported by grants-in-aid from MEXT/JSPS of Japan [Grant Nos. JP15H05886, JP21K03440, and JP24K00574], and Grant-in-Aid for Transformative Research Areas (A) ``Asymmetric Quantum Matters'', JSPS KAKENHI Grant No. JP23H04870, and “The Natural Laws of Extreme Universe”, Grant No. JP21H05191.
TO and DTA would like to thank the Royal Society of London for International Exchange funding, and DTA and DK thank EPSRC-UK (Grant No. EP/W00562X/1).


\bibliography{YbCuS2_ND_submittedNotes}
\bibliographystyle{apsrev4-1}

\end{document}